\documentclass[useAMS,usenatbib]{mn2e}
\usepackage{lmacs}
\usepackage{graphicx}
\usepackage{subfig}
\usepackage[T1]{fontenc} 
\usepackage{aecompl}
\usepackage{aas_macros}
\usepackage{color, soul}
\usepackage{gensymb}
\usepackage{amsmath}
\usepackage{url}
\graphicspath{{./figs/}}
\def\altaffilmark#1{$^{#1}$}
\def\altaffiltext#1#2{$^{#1}$#2}
\newcounter{aaffilcoun}
\newcommand\theaaffil{\addtocounter{aaffilcoun}{1}\theaaffilcoun}
\newcounter{affilcoun}
\newcommand\theaffil{\addtocounter{affilcoun}{1}\theaffilcoun}
\usepackage{xargs}                      

\title[Bimodality in Nebular Phase SNe~Ia]{Signatures of Bimodality in Nebular Phase Type Ia Supernova Spectra}

\author[Vallely et al.]
{P. J. Vallely\altaffilmark{\theaaffil},
M. A. Tucker\altaffilmark{\theaaffil},
B. J. Shappee\altaffilmark{2},
J. S. Brown\altaffilmark{1,\theaaffil},
K. Z. Stanek\altaffilmark{1,\theaaffil},
\newauthor
C. S. Kochanek\altaffilmark{1,4}
\\
\altaffiltext{\theaffil}{Department of Astronomy, The Ohio State University, 140 West 18th Avenue, Columbus, OH 43210, USA} \\
\altaffiltext{\theaffil}{Institute for Astronomy, University of Hawai'i, 2680 Woodlawn Drive, Honolulu, HI 96822, USA} \\
\altaffiltext{\theaffil}{Department of Astronomy and Astrophysics, University of California, Santa Cruz, CA 95064, USA} \\
\altaffiltext{\theaffil}{Center for Cosmology and AstroParticle Physics, The Ohio State University, 191 W. Woodruff Ave., Columbus, OH 43210, USA} \\
}

\begin{document}

\date{Accepted xxx Received xx; in original form xxx}

\pagerange{\pageref{firstpage}--\pageref{lastpage}} \pubyear{2019}

\maketitle

\label{firstpage}

\begin{abstract}
One observational prediction for Type Ia supernova (SNe~Ia) explosions produced through white dwarf--white dwarf collisions is the presence of bimodal velocity distributions for the $^{56}$Ni decay products, although this signature can also be produced by an off-center ignition in a delayed detonation explosion. These bimodal velocity distributions can manifest as double-peaked or flat-topped spectral features in late-time spectroscopic observations for favorable viewing angles. We present nebular-phase spectroscopic observations of 17 SNe~Ia obtained with the Large Binocular Telescope (LBT). Combining these observations with an extensive search of publicly available archival data, we collect a total sample of 48 SNe~Ia and classify them based on whether they show compelling evidence for bimodal velocity profiles in three features associated with $^{56}$Ni decay products: the [Fe~II] and [Fe~III] feature at $\sim5300$~{\AA}, the [Co~III] $\lambda5891$ feature, and the [Co~III] and [Fe~II] feature at $\sim6600$~{\AA}. We identify 9 bimodal SNe in our sample, and we find that these SNe have average peak $M_V$ about 0.3 mag fainter than those which do not. This is consistent with theoretical predictions for explosions created by nearly head-on collisions of white dwarfs due to viewing angle effects and $^{56}$Ni yields.
\end{abstract}

\section{Introduction}
Type Ia supernovae (SNe~Ia) are important objects in astronomy. With luminosities of $\sim10^{43}$ erg s$^{-1}$ at maximum light, they can be detected and monitored out to considerable distances. SNe~Ia are best known for their use as cosmological standardizable candles, arising from the tight correlation discovered by \cite{1993Phillips} between their peak $M_B$ and their rate of decline $\Delta m_{15}(B)$. \citet{1998Riess} and \citet{1999Perlmutter} took advantage of this relationship to discover the accelerating expansion of the universe. Beyond cosmology, SNe~Ia also play an important role in our understanding of nucleosynthesis, as they are one of the primary sources of iron-group and intermediate-mass elements, have a significant impact on the gas dynamics and star formation characteristics of galaxies, and are likely sources of high energy cosmic rays (see, e.g., \citealp{2014Maoz}).

Despite the importance of SNe~Ia, our knowledge of the events themselves is still far from complete. The most pressing questions surround the nature of their progenitors and explosion mechanism. SN~Ia explosions are the thermonuclear detonations of carbon-oxygen white dwarfs (CO WDs; \citealp{1960Hoyle,1969Colgate}), and a companion is required to trigger the explosion. The details of the explosion mechanism are unknown and remain an active topic of discussion. Possible progenitor scenarios can be broadly divided into two channels: one involving a companion star still undergoing thermonuclear burning (the single-degenerate or SD scenario), and one involving a WD companion (the double-degenerate or DD scenario).

In the canonical SD scenario a CO WD accretes hydrogen-rich or helium-rich material from a non-degenerate companion until it approaches the Chandrasekhar limit, at which point it experiences a thermonuclear runaway and explodes \citep{1973Whelan,2004Han}. There has also been considerable work done to study possible sub-Chandrasekhar \citep{WoosleyWeaver1994,2010Sim,2014Shen} and super-Chandrasekhar \citep{2004Yoon,2005Yoon,2012Hachisu} SD channel explosions. In the standard DD scenario, a tight WD binary loses energy and angular momentum to gravitational wave emission before undergoing tidal interactions and subsequently exploding as a SN~Ia \citep{1984Iben,1984Webbink,2012Shen}. The complete theoretical landscape for SNe~Ia is considerably more complex, including numerous proposed explosion mechanisms for both scenarios. Popular mechanisms include the delayed detonation \citep{Khokhlov1991,WoosleyWeaver1994,1999Livne} and double detonation \citep{1980Woosley,1982Nomoto,2007Bildsten,2014Shen}  models. More exotic mechanisms like the violent prompt merger scenario, a SD variant where a WD merges with the degenerate core of an asymptotic giant branch (AGB) star, have also been considered \citep{2003Livio,2013Soker}.

All of these progenitor channels have varying degrees of theoretical and observational problems. For instance, most SD scenario channels require finely tuned accretion rates in order for the WD to successfully gain mass and explode \citep{1972Starrfield,1982Nomoto,1984Iben}. Additionally, observational evidence for such progenitor systems has proven to be elusive. The nearby SNe~Ia 2011fe and 2014J were particularly well-studied \citep{2012Brown,2013Munari,2014Mazzali,2014Foley,2014Goobar,2016Galbany,2016Vallely,2017Shappee,2018Yang}, but no compelling evidence was found for the existence of non-degenerate companions \citep{2012Bloom,2012Chomiuk,2013Shappee,2014Margutti,2015Lundqvist}. 

Extensive searches for hydrogen emission lines at late times as evidence for stripped companion material have largely failed \citep{2005Mattila,2007Leonard,2013Lundqvist,2019Holmbo,2017Graham,2018Sand,2019bTucker,2016Maguire,2019Sand}. Indeed, in an unparalleled sample of over 100 SNe~Ia, \cite{2019Tucker} found no evidence for the predicted emission signatures. To date, only two normal Type~Ia SNe, ASASSN-18tb \citep{2018BrimacombeATel} and ATLAS18qtd \citep{2019Prieto}, show compelling evidence for strong H$\alpha$ emission \citep{2019Kollmeier}. However, \cite{2019Vallely} showed that the hydrogen signature in ASASSN-18tb is likely a product of CSM interaction and not indicative of a single-degenerate progenitor system. In contrast, \cite{2019Prieto} find that the H$\alpha$ emission observed in ATLAS18qtd is broadly consistent with the signatures expected for stripped companion material, although they note that the inferred hydrogen mass of $\sim 10^{-3} M_\odot$ is significantly below classical single-degenerate theoretical model predictions.

Fine tuning is also generally required for DD scenario mergers to avoid off-center ignitions and accretion-induced collapse to a neutron star \citep{1985Nomoto,2012Shen,2014Moll}.
Extensive discussion of SNe~Ia progenitor systems and explosion mechanisms and their respective theoretical and observational challenges can be found in \cite{2013Hillebrandt}, \cite{2014Maoz}, \cite{BranchWheeler2017}, and \cite{Ashall2018}.

Another possible progenitor scenario is the collisional WD channel. In this variant of the DD scenario, rather than slowly inspiralling due to gravitational wave emission, the two WDs collide nearly head-on -- virtually guaranteeing explosion due to the strong shocks produced in the collision \citep{2012Hawley,2013Kushnir,GS2013}. This scenario was first raised as a potential explanation for a small fraction of observed SNe~Ia in dense stellar regions \citep{2009Rosswog,2009Raskin,2010Raskin}. The \citeauthor{1962Kozai}-\citeauthor{1962Lidov} effect in triple systems may make this channel relatively generic \citep{2011Thompson,2014Antognini}. \cite{2012Katz} argue that the rate of direct WD collisions may nearly equal that of observed SNe~Ia, although the extent of this collision rate enhancement is debated by \cite{2013Hamers} and \cite{2018Toonen}.

The collisional WD channel provides fairly straightforward observable predictions. In particular, the velocity distribution of the $^{56}$Ni deposited in the ejecta of these explosions is intrinsically bimodal \citep{2015Dong}. At appropriate viewing angles, these bimodal velocity distributions will manifest as double-peaked or flat-topped spectral features in late-time spectroscopic observations of $^{56}$Ni decay products. Upon examining archival nebular phase spectra of SNe~Ia, \cite{2015Dong} confidently identified signatures of bimodality in 3 of the 18 SNe in their sample, indicating that SNe~Ia exhibiting this predicted characteristic are not uncommon.

This is not a unique observable of the collisional WD channel however, as bimodal $^{56}$Ni distributions can also be produced by an off-center delayed-detonation \citep{2007Fesen,2007Gerardy}. In this explosion mechanism, the supernova explosion begins as a sub-sonic deflagration wave at the center of the WD and propagates outward. The deflagration transitions into a supersonic detonation front when the density at its leading edge crosses a critical transition density\citep{Khokhlov1991,WoosleyWeaver1994,1999Livne}. This transition is not perfectly understood, so the value of this critical density is chosen such that the model replicates observed characteristics of SNe~Ia \citep{1995Hoeflich,2003Hoeflich}. A significant quantity of off-center $^{56}$Ni is produced during the detonation phase \citep{2002Hoeflich,2007Gerardy}. Off-center delayed detonations are generally considered in the context of SD progenitor systems, but they can also occur in the DD case \citep{2003Piersanti}.
The degeneracy between WD collisions and off-center delayed-detonations can be broken by using detailed radiative transfer calculations to analyze the observations \citep{2018Mazzali}.

In this paper, we present nebular-phase spectroscopic observations of 17 nearby SNe~Ia obtained over the past few years using the Large Binocular Telescope \citep[LBT;][]{2006HillLBT}. Most of these spectra were obtained as part of a long-term effort to accumulate a complete volume-limited spectroscopic sample of SNe~Ia nebular phase observations out to $z\sim0.2$. Once complete, the nebular spectra for 100 type IA Supernovae \citep[100IAs;][]{2018Dong100Ias} survey will be an invaluable resource for our understanding of SNe~Ia and their progenitors. Among the sample we present here, we identify 2 events that are consistent with an underlying bimodal velocity distribution. We then combine these spectra with a sample of 31 additional archival nebular phase SNe~Ia observations presented in \cite{2019Tucker}, where we identify an additional 7 events showing evidence of bimodality. We show that these bimodal SNe~Ia are systematically less luminous at peak than their single velocity component counterparts, and we discuss how this may arise from viewing angle dependent effects inherent to the collisional WD scenario or $^{56}$Ni production.

This paper is organized as follows. In Section~\ref{sec:obs} we describe the LBT observations we undertook to obtain 18 spectra of 17 nearby SNe~Ia during the nebular phase. In Section~\ref{sec:archival} we describe the sources from which we obtained our archival sample of nebular phase spectra and near-peak photometry, and we also provide a brief description of the methods we used to convert the observed $V$-band observations into absolute magnitudes. In Section~\ref{sec:bimodality} we describe the classification methods we use to determine whether or not spectra show evidence of a bimodal velocity distribution. Finally, in Section~\ref{sec:conclusion} we demonstrate that bimodal SNe~Ia are less luminous than SNe~Ia in general, and we discuss our findings in the context of the SNe~Ia progenitor problem.

\begin{figure*}
\centering
\includegraphics[scale=0.65]{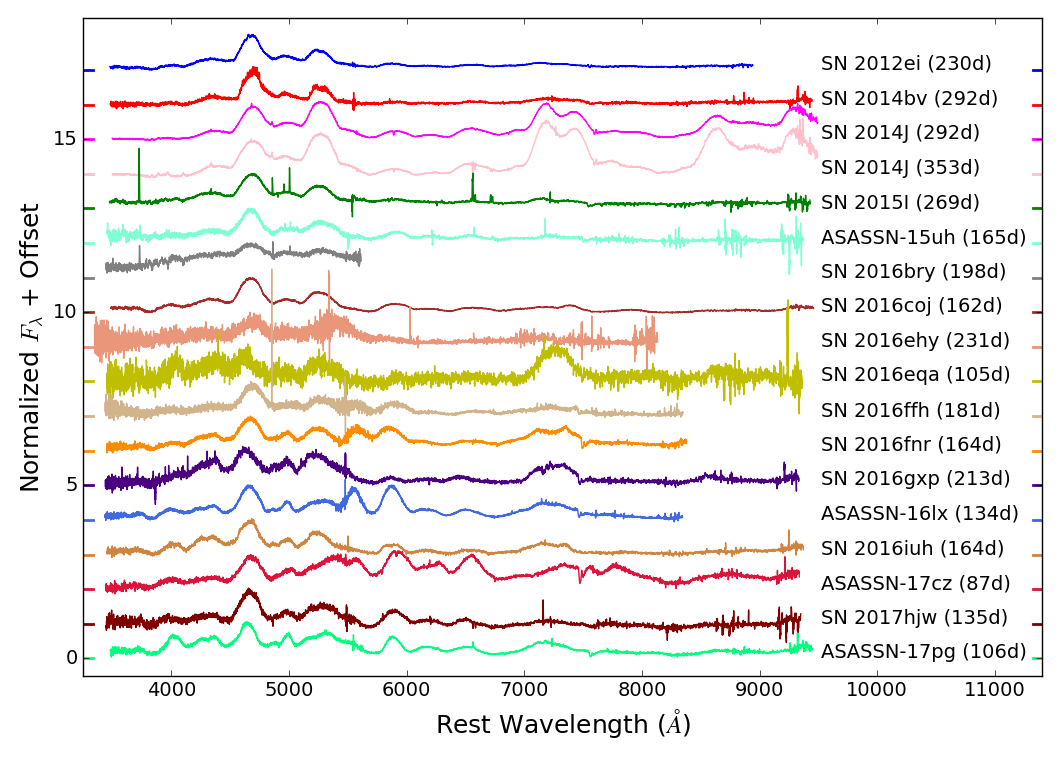}
\caption{The new LBT late-time SNe~Ia spectra we present in this paper. Phases relative to maximum $V$-band brightness are indicated in parentheses next to each spectrum. The colored tick marks on the vertical axes indicate the offset used when plotting each spectrum, and all spectra are normalized to the peak of the [Fe~III] $\lambda4701$ emission feature. This feature is prominent in all of the spectra, as are numerous other signatures of $^{56}$Ni decay products. We restrict our analysis to spectra with S/N$> 10$ and coverage of at least two of the pertinent Fe/Co features.}
\label{fig:LBTSpectra}
\end{figure*}

\section{The Sample}
\subsection{Previously Unpublished Observations}
\label{sec:obs}

All of the new spectra we present here were obtained using the Multi-Object Double Spectrographs mounted on the twin 8.4m Large Binocular Telescope \citep[MODS1 and MODS2;][]{PoggeMODS}. The MODS1 spectra were reduced using a combination of the \textsc{modsccdred}\footnote{\url{http://www.astronomy.ohio-state.edu/MODS/Software/modsCCDRed/}} \textsc{python} package, and the \textsc{modsidl} pipeline\footnote{\url{http://www.astronomy.ohio-state.edu/MODS/Software/modsIDL/}}. Unfortunately, some of the calibration data necessary to use the \textsc{modsidl} pipeline are not yet available for MODS2, so the MODS2 observations were reduced using standard techniques in IRAF to extract and calibrate the 1D spectra in wavelength and flux. Spectra of SNe 2016ehy, 2016ffh, 2016fnr, and ASASSN-16lx were obtained using only MODS2 data because MODS1 was not operational during those observations. All other spectra were obtained using only MODS1 data. Due to the relatively high sky noise in the red channel, the spectrum of 2016bry could only be extracted in the blue channel and is excluded from futher analysis.

The properties of these spectroscopic observations are summarized in Table~\ref{tab:spectra}, and Figure~\ref{fig:LBTSpectra} shows the 18 LBT spectra we obtained for this paper. Broadly speaking, the spectroscopic properties of our sample are comparable to the sample presented by \cite{2017Graham}. All of the spectra show prominent emission features in various $^{56}$Ni decay products, and the [Fe~III] $\lambda4701$ emission feature is particularly strong in all of the spectra. In our analysis, we focus on three neighboring Fe/Co emission features: the [Fe II] and [Fe III] blended feature at $\sim5300$~{\AA}, the [Co~III] $\lambda5891$ feature, and the [Fe II] and [Co~III] blended feature at $\sim6600$~{\AA}.

An in-depth discussion justifying the use of these features can be found in Appendix B of \cite{2015Dong}. In short, they are chosen because they are narrow, well-characterized, and nearly identical between the spectra of SNe~1991bg and SN 1999by. The structure of the [Fe~III] $\lambda4701$ feature, on the other hand, differs significantly between the two SNe and appears to be strongly impacted by complicated blends of nearby lines, rendering it unsuitable for our analysis. We are limited to these optical features only due to the wavelength ranges of the spectra in our sample. In principle, an underlying bimodal velocity distribution of $^{56}$Ni should manifest in all late time Fe/Co features, and as we discuss in Section~\ref{sec:bimodality}, this allows us to verify our identifications using the results of studies at non-optical wavelengths.

We restrict our analysis to spectra with S/N$> 10$ and coverage of at least two of the pertinent Fe/Co features. This leads to the exclusion of the spectra for SNe 2016bry, 2016ehy, and 2016eqa. As we discuss further in Section~\ref{sec:bimodality}, the spectra of SNe 2014bv and 2016iuh are particularly interesting, as they show fairly compelling evidence of bimodal velocity distributions in their $^{56}$Ni ejecta.

\begin{table*}
\caption{LBT Spectroscopic Observations.}
\label{tab:spectra}
\begin{tabular}{llrcrll}
\hline\hline
SN & Obs. Date & Phase (d) & z & Exposure (s) &  Wavelength Coverage & bimodal Fe/Co\\
\hline\hline
SN 2012ei & 2013-05-03 & 230 & 0.00657 & 4800 & 3470 -- 8940 {\AA}  & No \\ 
SN 2014bv & 2015-04-22 & 292 & 0.00559 & 10800 & 3480 -- 9440 {\AA}  & \textbf{Yes} \\ 
SN 2014J & 2014-11-21 & 292 & 0.00068 & 12600 & 3490 -- 9490 {\AA}  & No \\ 
SN 2014J & 2015-01-21 & 353 & 0.00068 & 10800 & 3490 -- 9490 {\AA}  & No \\ 
SN 2015I & 2016-02-08 & 269 & 0.00759 & 10800 & 3470 -- 9420 {\AA}  & No \\ 
ASASSN-15uh & 2016-06-14 & 165 & 0.01350 & 3600 & 3450 -- 9370 {\AA}  & No \\
SN 2016bry$^b$ & 2016-11-20 & 198 & 0.01602 & 2700 & 3440 -- 5600 {\AA}  & Noisy \\
SN 2016coj & 2016-11-18 & 162 & 0.00448 & 2700 & 3480 -- 9450 {\AA}  & No \\
SN 2016ehy$^a$ & 2017-03-02 & 231 & 0.04500 & 3600 & 3340 -- 8130 {\AA}  & Noisy \\
SN 2016eqa & 2016-11-20 & 105 & 0.01496 & 5400 & 3440 -- 9360 {\AA}  & Noisy \\
SN 2016ffh$^a$ & 2017-03-02 & 181 & 0.01820 & 3600 & 3430 -- 8340 {\AA}  & No \\
SN 2016fnr$^a$ & 2017-03-02 & 164 & 0.01437 & 3600 & 3450 -- 8370 {\AA}  & No \\
SN 2016gxp & 2017-05-27 & 213 & 0.01785 & 3600 & 3430 -- 9330 {\AA}  & No \\
ASASSN-16lx$^a$ & 2017-03-02 & 134 & 0.01860 & 3600 & 3430 -- 8340 {\AA}  & No \\
SN 2016iuh & 2017-05-27 & 164 & 0.01370 & 7200 & 3450 -- 9370 {\AA}  & \textbf{Yes [Co~III]} \\
ASASSN-17cz & 2017-05-28 & 87 & 0.01738 & 2700 & 3440 -- 9330 {\AA}  & No \\
SN 2017hjw & 2018-03-14 & 135 & 0.01616 & 2700 & 3440 -- 9340 {\AA}  & No \\ 
ASASSN-17pg & 2018-03-14 & 106 & 0.00562 & 1800 & 3480 -- 9440 {\AA}  & No \\
\hline\hline
\end{tabular} \\
\begin{flushleft}
The wavelength coverage is reported for the rest frame of each spectrum. \\
$^a$MODS2 spectra. \\
$^b$The red channel spectrum had too low a S/N to effectively extract from the observations.
\end{flushleft}
\end{table*}

\subsection{Archival Data}
\label{sec:archival}
We also utilize a subset of the exhaustive sample of spectroscopic archival late-time SNe~Ia observations we collected and present in \cite{2019Tucker}. To obtain this sample we systematically extracted spectra from a number of archival databases, including the Weizmann Interactive Supernova Data Repository \citep[WISeREP;][]{2012WISeREP}, the Open Supernova Catalog \citep[OSC;][]{2017OSC}, the Berkeley SuperNova Ia Program \citep[BSNIP;][]{2009BSNIP,2012Silverman}, the Carnegie Supernova Project \citep[CSP;][]{2013Folatelli}, and the Center for Astrophysics Supernova Program \citep{2012Blondin}. We also present a number of previously unpublished spectra reduced from raw data available in other public archives. See \cite{2019Tucker} for a detailed description of how we obtained the full sample. The spectroscopic properties of the SNe we utilize in our analysis are summarized in Table~\ref{tab:archivalspec}.

We restrict our sample to events with reasonably well-sampled near-peak $V$-band light curves, so that we can compare the peak luminosities of the bimodal events with the overall sample. The photometric properties of these archival SNe~Ia are summarized in Table~\ref{tab:archivalphot}. Only \textit{Swift} UVOT observations were obtained for a few of these events, and in these instances we use UVOT $V$-band observations in lieu of Johnson $V$-band observations. We identify these events in Table~\ref{tab:archivalphot}. In cases where both were obtained, we use the Johnson $V$ observations.

With the exception of the extreme cases of SNe 1986G, 2002er, and 2014J, we do not correct for host galaxy extinction. It seems unlikely that there would be any preference for bimodal events to occur along high or low extinction lines of sight, so for our relative comparison of peak $M_V$ this should be of no major concern. We account for Galactic foreground extinction using the \cite{2011Schlafly} infrared-based dust map, and we use redshift-independent estimates of the distance modulus $(\mu)$ for SNe with $z<0.01$.

\begin{table*}
\caption{Spectroscopic Properties}
\label{tab:archivalspec}
\begin{tabular}{llllll}
\hline\hline
SN & z & Type & Phase (d) & bimodal Fe/Co & Reference(s) \\
\hline\hline
SN 2012cg & 0.00146 & Ia-Norm & 284 & No & \cite{2018Shappee} \\
SN 2014bv & 0.00559 & Ia-Norm & 292 & \textbf{Yes} & This Work \\
SN 2014J & 0.00068 & Ia-Norm & 292,353 & No & This Work \\
SN 2015I & 0.00759 & Ia-Norm & 269 & No & This Work \\
SN 2016coj & 0.00448 & Ia-Norm & 162 & No & This Work \\
SN 2016fnr & 0.01437 & Ia-Norm & 164 & No & This Work \\
SN 2016gxp & 0.01785 & Ia-91T & 213 & No & This Work \\
SN 2016iuh & 0.01370 & Ia-91bg & 164 & \textbf{Yes [Co~III]} & This Work \\
SN 2017hjw & 0.01616 & Ia-Norm & 135 & No & This Work \\
ASASSN-15uh & 0.01350 & Ia-91T & 165 & No & This Work \\
ASASSN-16lx & 0.01860 & Ia-Norm & 134 & No & This Work \\
ASASSN-17cz & 0.01738 & Ia-Norm & 87 & No & This Work \\
ASASSN-17pg & 0.00562 & Ia-Norm & 106 & No & This Work \\
ASASSN-14jg & 0.01483 & Ia-Norm & 216 & No & \cite{2019Tucker} \\
SN 1981B & 0.00603 & Ia-Norm & 113,267 & No & \cite{1983Branch,2001SUSPECT} \\
SN 1986G & 0.00180 & Ia-91bg & 256 & \textbf{Yes} & \cite{1992Ruiz} \\
SN 1989B & 0.00243 & Ia-Norm & 150 & No & \cite{1994Wells} \\
SN 1990N & 0.00340 & Ia-Norm & 184,225,253,278 & No & \cite{1998Gomez} \\
SN 1991T & 0.00579 & Ia-91T & 183$^1$,255$^2$,281$^2$,313$^2$,317$^1$ & No & $^1$\cite{2012Silverman};$^2$\cite{1998Gomez} \\
SN 1998aq & 0.00370 & Ia-Norm & 230,240 & No & \cite{2012Blondin} \\
SN 1998bu & 0.00299 & Ia-Norm & 190$^1$,208$^1$,217$^1$,236$^2$,243$^1$, & No & $^1$\cite{2012Blondin};$^2$\cite{2012Silverman}; \\
  &  &  & 280$^2$,329$^3$ &  & $^3$\cite{2001Cappellaro} \\
SN 1999aa & 0.01444 & Ia-91T & 256 & No & \cite{2012Silverman} \\
SN 1999by & 0.00213 & Ia-91bg & 181 & No & \cite{2012Silverman} \\
SN 2000cx & 0.00793 & Ia-Pec & 181 & No & \cite{2012Blondin} \\
SN 2002dj & 0.00939 & Ia-Norm & 218,271 & No & \cite{2008Pignata} \\
SN 2002er & 0.00857 & Ia-Norm & 214 & \textbf{Yes [Co~III]} & \cite{2005Kotak} \\
SN 2003du & 0.00638 & Ia-Norm & 219 & No & \cite{2007Stanishev} \\
SN 2003gs & 0.00477 & Ia-Pec & 197 & \textbf{Yes} & \cite{2012Silverman} \\
SN 2003hv & 0.00560 & Ia-Norm & 319 & \textbf{Yes} & \cite{2009Leloudas} \\
SN 2004bv & 0.01061 & Ia-Norm & 159 & No & \cite{2012Silverman} \\
SN 2004eo & 0.01570 & Ia-Norm & 226 & No & \cite{2007Pastorello04eo} \\
SN 2005am & 0.00790 & Ia-Norm & 297,380 & \textbf{Yes} & \cite{2007Leonard} \\
SN 2005cf & 0.00646 & Ia-Norm & 264 & No & \cite{2007Leonard} \\
SN 2007af & 0.00546 & Ia-Norm & 301 & No & \cite{2012Blondin} \\
SN 2007le & 0.00672 & Ia-Norm & 304 & No & \cite{2012Silverman} \\
SN 2007on & 0.00649 & Ia-Norm & 284 & \textbf{Yes} & \cite{2013Folatelli} \\
SN 2008A & 0.01646 & Ia-02cx & 201,225 & No & \cite{2014McCully} \\
SN 2008Q & 0.00794 & Ia-Norm & 201 & No & \cite{2012Silverman} \\
SN 2011by & 0.00284 & Ia-Norm & 204,308 & No & \cite{2013Silverman} \\
SN 2011fe & 0.00080 & Ia-Norm & 205$^1$,226$^1$,229$^2$,259$^1$,347$^1$ & No & $^1$\cite{2015Mazzali};$^2$\cite{2013Shappee} \\
SN 2011iv & 0.00649 & Ia-Norm & 244,261 & \textbf{Yes} & \cite{2018Gall} \\
SN 2012fr & 0.00540 & Ia-Norm & 220,259,338,365 & No & \cite{2015Childress} \\
SN 2012hr & 0.00756 & Ia-Norm & 281 & No & \cite{2015Childress} \\
SN 2013aa & 0.00400 & Ia-Norm & 188,205,345 & No & \cite{2015Childress} \\
SN 2013dy & 0.00389 & Ia-Norm & 332 & No & \cite{2015Pan} \\
SN 2013gy & 0.01402 & Ia-Norm & 271 & No & \cite{2015Childress} \\
SN 2015F & 0.00489 & Ia-Norm & 194,293 & No & \cite{2019Tucker} \\
SN 2017cbv & 0.00400 & Ia-Norm & 315 & No & \cite{2019Tucker} \\
\hline\hline
\end{tabular} \\
\begin{flushleft}
Note that phases are calculated in the observed frame relative to maximum $V$-band brightness.
\end{flushleft}
\end{table*}

\begin{table*}
\caption{Peak $V$-band Brightness}
\label{tab:archivalphot}
\begin{tabular}{llllll}
\hline\hline
SN & $m_V$ & Dist. Mod. $(\mu)$ & Extinction $(A_V)$ & $M_V$ &  Reference(s) \\
\hline\hline
SN 2012cg & 11.90$^1$ & 31.02$^2$ & 0.057 & $-$19.18 & $^1$\cite{2018Vinko}; $^2$\cite{2013Munari} \\
SN 2014bv$^a$ & 13.92$^1$ & 32.17$^2$ & 0.106 & $-$18.36 & $^1$\cite{2014SOUSA}; $^2$\cite{2013Tully} \\
SN 2014J & 10.56$^1$ & 27.64$^2$ & 0.435+1.76$^b$ & $-$19.28 & $^1$\cite{2014Tsvetkov};  \\
& & & & & $^2$\cite{2009Dalcanton}; $^2$\citet{2015Marion} \\
SN 2015I$^a$ & 13.99$^1$ & 32.64$^2$ & 0.182 & $-$18.83 & $^1$\cite{2014SOUSA}; $^2$\cite{2013Tully}  \\
SN 2016coj & 13.02$^1$ & 31.90$^2$ & 0.052 & $-$18.93 & $^1$\cite{2017ASASSNLightCurves}; $^2$\cite{2001Blakeslee}  \\
SN 2016fnr & 15.28$^1$ & 33.98 & 0.128 & $-$18.83 & $^1$\cite{2017ASASSNLightCurves}  \\
SN 2016gxp & 14.84$^1$ & 34.55 & 0.338 & $-$20.05 & $^1$\citeauthor{2019Chen} (2019, in prep)  \\
SN 2016iuh & 15.43$^1$ & 33.88 & 0.045 & $-$18.49 & $^1$\cite{2017ASASSNLightCurves}  \\
SN 2017hjw & 15.85$^1$ & 34.24 & 0.370 & $-$18.76 & $^1$\citeauthor{2019Chen} (2019, in prep)  \\
ASASSN-15uh & 15.28$^1$ & 33.85 & 0.410 & $-$18.98 & $^1$\cite{2017ASASSNLightCurves} \\
ASASSN-16lx & 15.47$^1$ & 34.55 & 0.115 & $-$19.20 & $^1$\cite{2017ASASSNLightCurves} \\
ASASSN-17cz & 16.60$^1$ & 34.33 & 1.138 & $-$18.86 & $^1$\citeauthor{2019Chen} (2019, in prep) \\
ASASSN-17pg & 14.46$^1$ & 32.51$^2$ & 0.145 & $-$18.19 & $^1$\cite{2017ASASSNLightCurves}; $^2$\cite{2016Tully} \\
ASASSN-14jg & 14.92$^1$ & 34.05 & 0.042 & $-$19.17 & $^1$\cite{2017ASASSNLightCurves} \\
SN 1981B & 11.85$^1$ & 30.83$^2$ & 0.050 & $-$19.03 & $^1$\cite{1982Barbon}; $^2$\cite{2013Tully} \\
SN 1986G & 11.44$^1$ & 27.82$^2$ & 1.95$^c$ & $-$18.33 & $^1$\cite{1987Phillips}; $^2$\cite{2013Tully} \\
SN 1989B & 11.99$^1$ & 29.78$^2$ & 0.091 & $-$17.88 & $^1$\cite{1994Wells}; $^2$\cite{2013Tully} \\
SN 1990N & 12.73$^1$ & 31.72$^2$ & 0.071 & $-$19.06 & $^1$\cite{1998Lira}; $^2$\cite{2013Tully} \\
SN 1991T & 11.51$^1$ & 30.91$^2$ & 0.060 & $-$20.00 & $^1$\cite{1998Lira}; $^2$\cite{2000Parodi} \\
SN 1998aq & 12.46$^1$ & 31.67$^2$ & 0.039 & $-$19.25 & $^1$\cite{2005Riess}; $^2$\cite{2013Tully} \\
SN 1998bu & 11.86$^1$ & 30.11$^2$ & 0.069 & $-$18.32 & $^1$\cite{1999Jha}; $^2$\cite{2013Tully} \\
SN 1999aa & 14.90$^1$ & 34.10$^1$ & 0.109 & $-$19.31 & $^{1}$\cite{2008Kowalski} \\
SN 1999by & 13.14$^1$ & 30.82$^2$ & 0.054 & $-$17.73 & $^1$\cite{2004Garnavich}; $^2$\cite{2013Tully} \\
SN 2000cx & 13.23$^1$ & 32.40$^2$ & 0.224 & $-$19.39 & $^1$\cite{2001Li}; $^2$\cite{2008Takanashi} \\
SN 2002dj & 14.13$^1$ & 32.65$^2$ & 0.261 & $-$18.78 & $^1$\cite{2009Hicken}; $^2$\cite{2013Tully} \\
SN 2002er & 14.59$^1$ & 32.51$^2$ & 1.12$^d$ & $-$19.04 & $^1$\cite{2004Pignata}; $^2$\cite{2013Tully} \\
SN 2003du & 13.57$^1$ & 32.83$^2$ & 0.027 & $-$19.29 & $^1$\cite{2009Hicken}; $^2$\cite{2013Tully} \\
SN 2003gs & 13.49$^1$ & 31.49$^2$ & 0.097 & $-$18.10 & $^1$\cite{2009Krisciunas}; $^2$\cite{2001Blakeslee} \\
SN 2003hv & 12.55$^1$ & 31.55$^2$ & 0.042 & $-$19.04 & $^1$\cite{2009Leloudas}; $^2$\cite{2013Tully} \\
SN 2004bv & 14.02$^1$ & 32.80$^2$ & 0.174 & $-$18.95 & $^1$\cite{2010Ganeshalingam}; $^2$\cite{2013Tully} \\
SN 2004eo & 15.33$^1$ & 34.12$^1$ & 0.296 & $-$19.09 & $^1$\cite{2007Pastorello04eo} \\
SN 2005am & 13.76$^1$ & 32.24$^2$ & 0.147 & $-$18.63 & $^1$\cite{2010Ganeshalingam}; $^2$\cite{2013Tully} \\
SN 2005cf & 13.50$^1$ & 32.32$^2$ & 0.267 & $-$19.09 & $^1$\cite{2007Pastorello05cf} \\
SN 2007af & 13.21$^1$ & 31.76$^2$ & 0.107 & $-$18.66 & $^1$\cite{2009Hicken}; $^2$\cite{2013Tully} \\
SN 2007le & 13.66$^1$ & 31.73$^2$ & 0.092 & $-$18.16 & $^1$\cite{2012Hicken}; $^2$\cite{2009Springob} \\
SN 2007on & 12.96$^1$ & 31.45$^2$ & 0.032 & $-$18.52 & $^1$\cite{2010ContrerasCSP}; $^2$\cite{2013Tully} \\
SN 2008A & 16.09$^1$ & 34.05 & 0.149 & $-$18.11 & $^1$\cite{2010Ganeshalingam} \\
SN 2008Q & 13.75$^1$ & 32.30 & 0.227 & $-$18.78 & $^1$\cite{2010Ganeshalingam} \\
SN 2011by$^a$ & 12.92$^1$ & 32.01$^2$ & 0.038 & $-$19.13 & $^1$\cite{2014SOUSA}; $^2$\cite{2012Maguire} \\
SN 2011fe & 9.97$^1$ & 29.05$^2$ & 0.024 & $-$19.10 & $^1$\cite{2013Munari}; $^2$\cite{2012Vinko} \\
SN 2011iv & 12.38$^1$ & 31.45$^2$ & 0.031 & $-$19.10 & $^1$\cite{2018Gall}; $^2$\cite{2013Tully} \\
SN 2012fr & 11.98$^1$ & 31.25$^2$ & 0.056 & $-$19.33 & $^1$\cite{2018Contreras}; $^2$\cite{2013Tully} \\
SN 2012hr$^a$ & 13.75$^1$ & 33.03$^2$ & 0.124 & $-$19.40 & $^1$\cite{2014SOUSA}; $^2$\cite{2013Tully} \\
SN 2013aa$^a$ & 11.62$^1$ & 30.55$^2$ & 0.466 & $-$19.40 & $^1$\cite{2014SOUSA}; $^2$\cite{1985Bottinelli} \\
SN 2013dy & 12.94$^1$ & 30.68$^2$ & 0.421 & $-$18.16 & $^1$\cite{2016Zhai}; $^2$\cite{2009Tully} \\
SN 2013gy & 14.77$^1$ & 33.75 & 0.158 & $-$19.14 & $^1$\cite{2017Graham} \\
SN 2015F & 13.27$^1$ & 31.64$^2$ & 0.556 & $-$18.93 & $^1$\cite{2017Graham}; $^2$\cite{2017Cartier} \\
SN 2017cbv & 11.64$^1$ & 30.13$^2$ & 0.463 & $-$18.95 & $^1$\citeauthor{2019Chen} (2019, in prep); $^2$\cite{1985Bottinelli} \\
\hline\hline
\end{tabular} \\
\begin{flushleft}
The foreground Galactic extinction $A_V$ values are taken from \cite{2011Schlafly}.\\
$^a$Johnson $V$ observations were not obtained, so \textit{Swift} UVOT $V$ observations are substituted. \\
$^b$SN 2014J exhibits considerable host galaxy extinction, so we adopt the $A_V=1.76$ host galaxy extinction value from \cite{2014Tsvetkov}. \\
$^c$SN 1986G exhibits considerable host galaxy extinction, so we adopt $E(B-V)=0.63$ from \cite{1987diSerego} and assume $R_V=3.1$ to obtain $A_V=1.95$. \\
$^d$SN 2002er exhibits considerable host galaxy extinction, so we adopt $E(B-V)=0.36$ from \cite{2004Pignata} and assume $R_V=3.1$ to obtain $A_V=1.12$.
\end{flushleft}
\end{table*}

\section{Determining bimodality}
\label{sec:bimodality}

\begin{figure}
\centering
\includegraphics[width=\columnwidth]{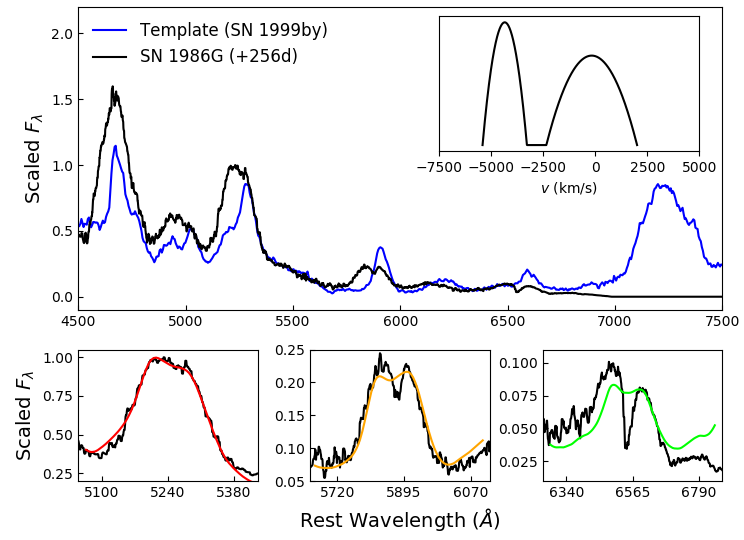}
\caption{An illustration of our convolution-based fitting technique. The nebular phase spectrum of SN 1986G is shown in black, and the SN 1999by spectrum we adopt as a pre-convolution template throughout the paper is shown in blue. The template spectrum is convolved with the velocity kernel plotted in the upper right panel to produce the feature-specific fits in the lower panels. As in all figures, the convolution fit to the $\sim5300$~{\AA} [Fe~II]/[Fe~III] feature is shown in red, the convolution fit to the [Co~III] $\lambda5891$ feature is shown in orange, and the convolution fit to the $\sim6600$~{\AA} [Co~III]/[Fe~II] feature is shown in green. Complications in the interpretation of the SN~1986G spectrum are discussed in Section~\ref{sec:bimodality}.}
\label{fig:Template}
\end{figure}

\begin{figure*}
\centering
\includegraphics[scale=0.4]{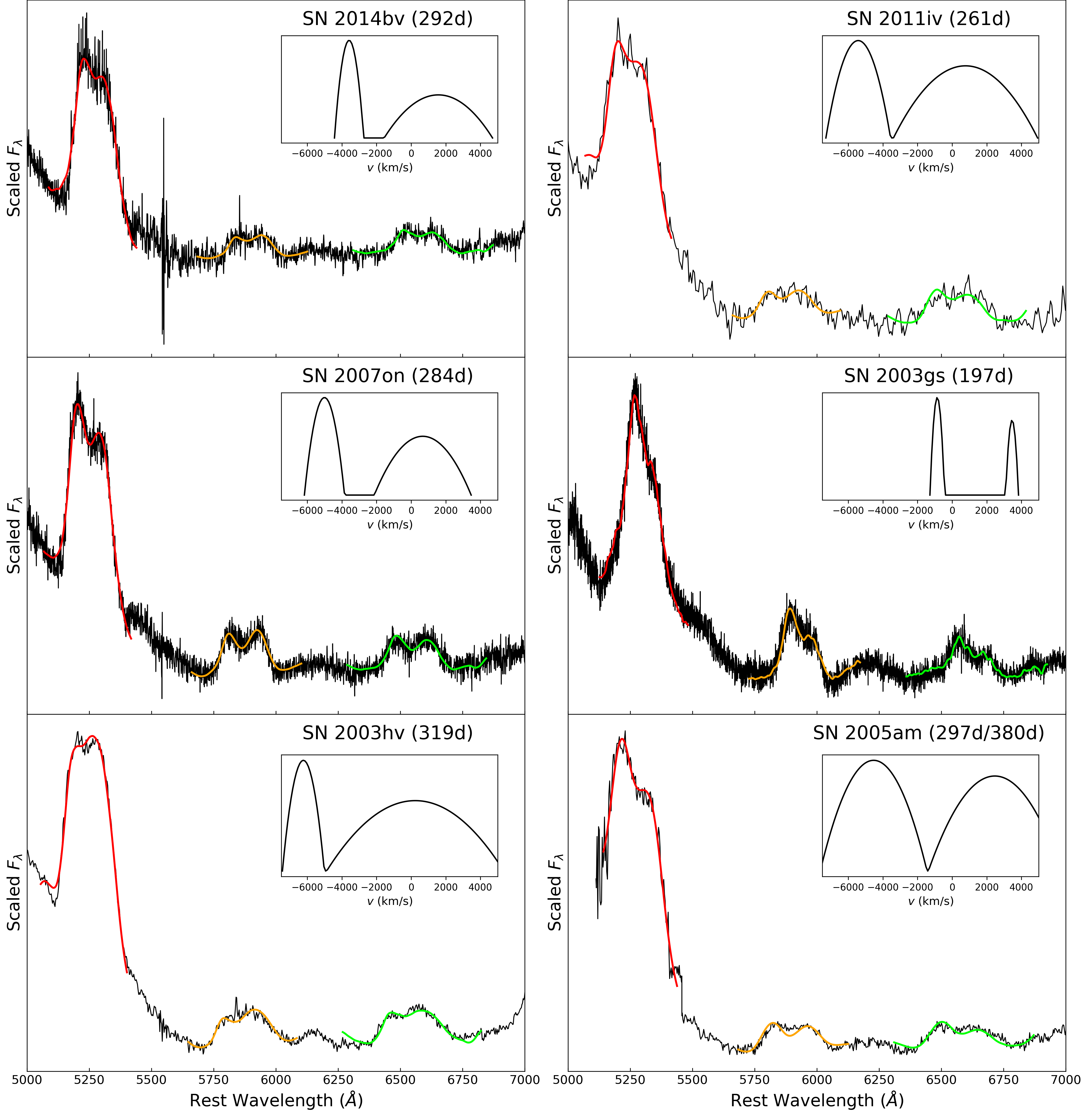}
\caption{The six spectra we confidently identify as bimodal through multi-feature fits to all three of the relevant $^{56}$Ni decay features. In all cases the two velocity kernel components are distinctly separated, and in most cases they are separated by a considerable margin. All of the fits appear reasonable upon inspection.}
\label{fig:MultiFit}
\end{figure*}

In the majority of cases, detecting signatures of bimodality in the $^{56}$Ni velocity profiles of a spectrum can be performed fairly reliably by inspection. For instance, one can visually identify double-peaked Fe/Co features in the late time spectra of SN~2007on and SN~2014bv, indicating possible bimodal velocity profiles. (see Figure~\ref{fig:MultiFit}). Similarly, a cursory examination of the late-time spectra of SN~2011fe and SN~2012cg shows no need to invoke anything beyond standard single-component velocity broadening. However, there are also events like SNe 2016iuh and 2012ei which do not fall cleanly into either category. In order to handle the classification of these borderline events self-consistently, and to minimize the impact of any potential bias, it is best to have an objective classification scheme.

We use the direct convolution technique described by \cite{2015Dong}, although our implementation differs slightly. We construct a bimodal velocity kernel using two quadratic components, and then convolve this kernel with a template SN~Ia nebular phase spectrum.
Due to its narrow emission features and particularly high S/N, we retain from \cite{2015Dong} the use of SN~1999by for this pre-convolution template. This phase +180d SN~1999by template spectrum is shown by the blue line in Figure~\ref{fig:Template}.

The velocity convolution kernel is described by

\begin{equation}
\frac{dM}{dv_{LOS}} \propto P_1 + r\cdot P_2 ,\text{ where}
\end{equation}

\begin{equation}
P_1 = \max \left( 1 - \frac{(v_{LOS}-v_{shift,1})^2}{\sigma_{mod,1}^2}, 0 \right),
\end{equation}

\begin{equation}
v_{shift,1} = v_{shift} - \frac{1}{2}v_{sep},
\end{equation}

\begin{equation}
P_2 = \max \left( 1 - \frac{(v_{LOS}-v_{shift,2})^2}{\sigma_{mod,2}^2}, 0 \right),\text{ and}
\end{equation}

\begin{equation}
v_{shift,2} = v_{shift} + \frac{1}{2}v_{sep}.
\end{equation}

There are 5 free parameters: the shifts of the two components $v_{shift,1}$ and $v_{shift,2}$, the widths of the two components $\sigma_{mod,1}$ and $\sigma_{mod,2}$, and the peak ratio of the components $r$. The shifts are described using the two parameters of velocity shift, $v_{shift} = \frac{1}{2} ( \sigma_{mod,1} + \sigma_{mod,2} )$, and velocity separation, $v_{sep} = v_{shift,2} - v_{shift,1}$.

\begin{figure*}
\centering
\includegraphics[scale=0.4]{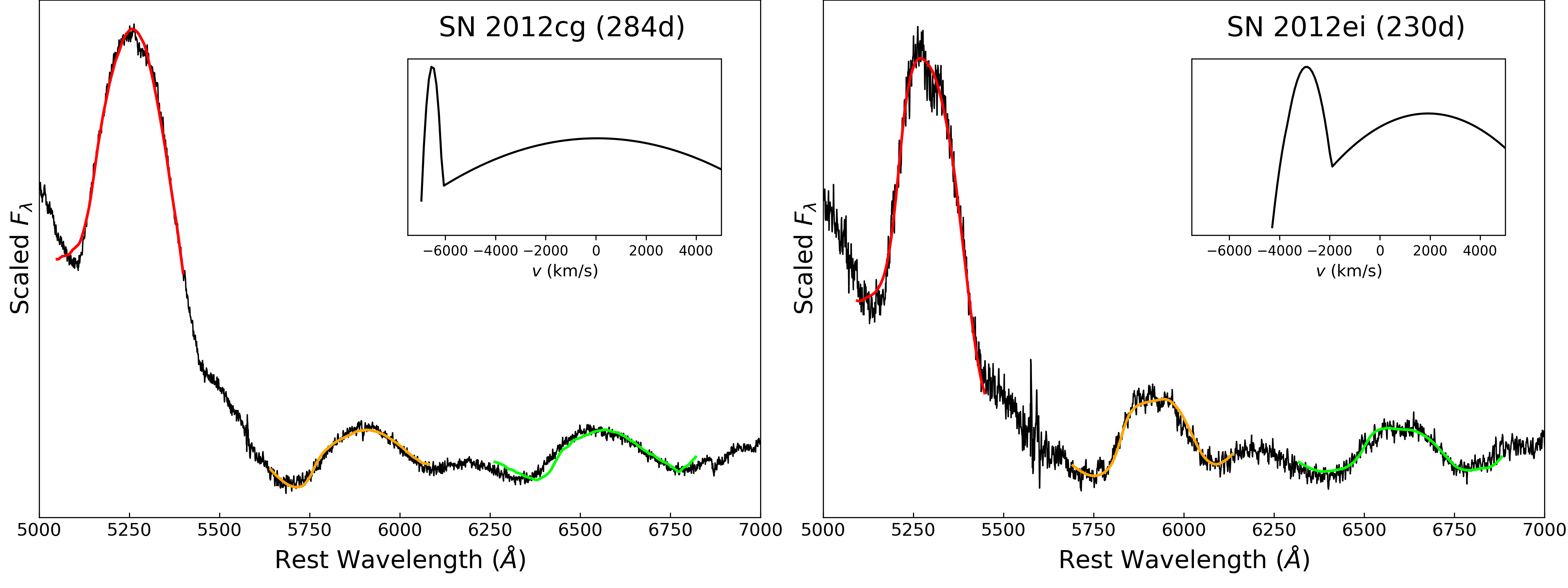}
\caption{Two spectra representative of the single-component events we identify in our sample: SNe 2012cg and 2012ei. Fitting a convolution kernel to SN~2012cg is somewhat unnecessary, as the spectrum clearly does not show signatures of bimodality. SN~2012ei, however, does show some signs of bimodality  with somewhat flat-topped emission features and a degree of double-peaked emission in the $\sim$5300~{\AA} [Fe~II] and [Fe~III] feature. This demonstrates the importance of using an objective classification scheme. In both cases, we classify these spectra as single-component because the best-fit velocity kernel has components that are significantly overlapping. Both spectra were obtained using the LBT, and the SN 2012cg spectrum was previously published by \protect\cite{2018Shappee}.}
\label{fig:SingleComponentFit}
\end{figure*}

\begin{figure*}
\centering
\includegraphics[scale=0.4]{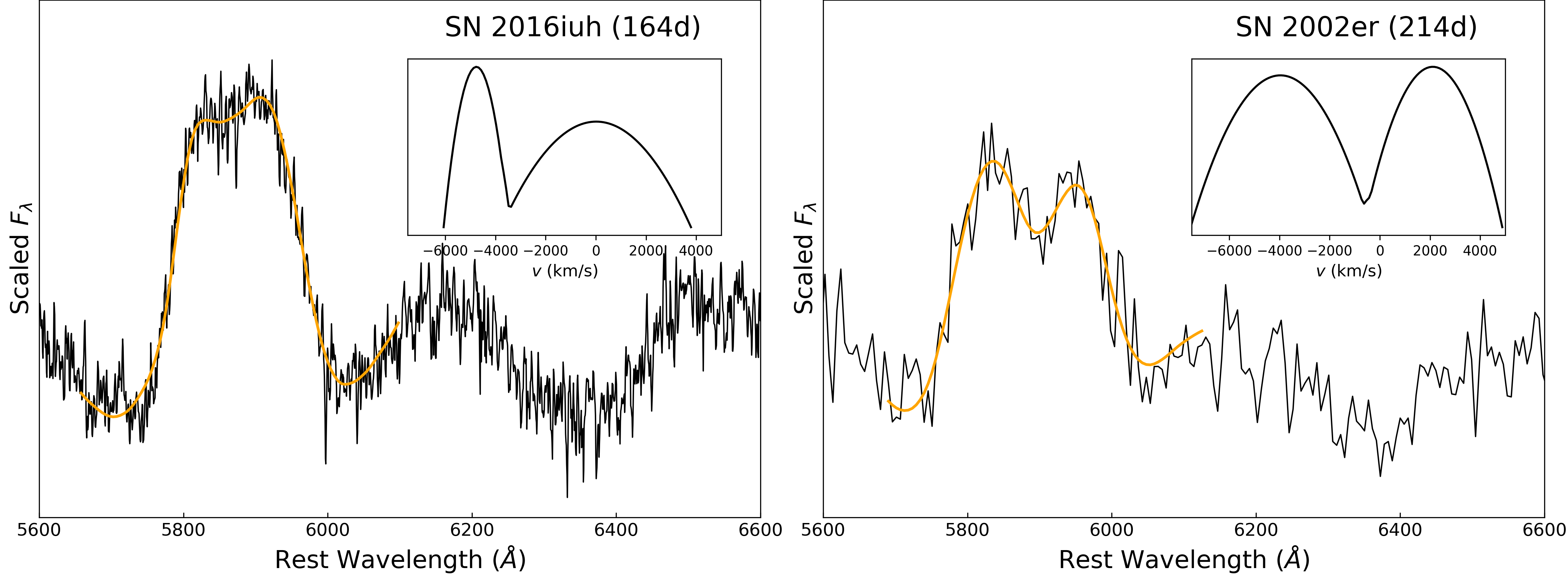}
\caption{The two additional spectra we tentatively identify as bimodal based only on fits to the [Co~III] $\lambda5891$ feature. While these identifications are less robust than those obtained using the multifeature fits, they appear to be consistent with the rest of the bimodal sample (see Section~\ref{sec:conclusion}).}
\label{fig:CoIIIFit}
\end{figure*}

We limit our analysis to three features we can confidently associate with $^{56}$Ni decay products: the [Fe~II] and [Fe~III] feature at $\sim5300$~{\AA}, the [Co~III] $\lambda5891$ feature, and the [Co~III] and [Fe~II] feature at $\sim6600$~{\AA}. The [Co~III] $\lambda5891$ feature is particularly valuable due to its lack of multi-line blending. We obtain fits for each spectrum by varying the velocity kernel parameters ($v_{shift}$, $\sigma_{mod,1}$, $\sigma_{mod,2}$,  $v_{sep}$, and $r$) to minimize $\chi^2$ for two cases -- one fit using only the [Co~III] $\lambda5891$ feature, and one fit using all three of the pertinent $^{56}$Ni decay features. The fit parameters of the spectra we show in Figures~\ref{fig:Template}-\ref{fig:CoIIIFit} are provided in Table~\ref{tab:fitparams}. In all figures, the convolution fit to the $\sim5300$~{\AA} [Fe~II]/[Fe~III] feature is shown in red, the convolution fit to the [Co~III] $\lambda5891$ feature is shown in orange, and the convolution fit to the $\sim6600$~{\AA} [Co~III]/[Fe~II] feature is shown in green.

We classify the spectral fits as being consistent with a bimodal velocity profile if the two quadratic components of the velocity kernel do not significantly overlap -- that is, $v_{sep} \gtrsim \sigma_{mod,1} + \sigma_{mod,2}$. While there are benefits to using only the [Co~III] feature -- namely that it is not subject to blending concerns -- we regard identifications made using all three features as more robust. It is very unlikely that a single-component velocity distribution can produce similarly spaced double-peaked profiles for the three well-separated features. For robust identification using the triple-feature fit we strictly require $v_{sep} > \sigma_{mod,1} + \sigma_{mod,2}$, while for the more tentative [Co~III] feature identifications we allow for a small overlap of $500$ km s$^{-1}$ provided that the two kernel components are still largely distinct from one another and satisfy $\sigma_{mod,1} < v_{sep}$ and $\sigma_{mod,2} < v_{sep}$ (see Figure~\ref{fig:CoIIIFit}).

In nearly all cases where the three-feature fit produces a bimodal classification, the single [Co~III] feature fit does so as well. Out of the six events classified as bimodal using the multi-feature fit, only in the case of SN~2003hv does the single [Co~III] feature fit disagree. Upon inspection of the SN~2003hv spectrum, one notes that the [Co~III] $\lambda5891$ feature, although not double-peaked, shows a flat-top profile consistent with a bimodal velocity profile. Our sample includes 14 of the 18 SNe considered by \cite{2015Dong}. We recover the bimodal classifications found by \cite{2015Dong} for SNe 2007on, 2003gs, and 2005am. Using our triple-feature fit criteria, we classify SNe 2008Q and 2003hv, which \cite{2015Dong} had described as ambiguous identifications, as single-component and bimodal events, respectively.

It is encouraging to note that the results of our classification scheme generally agree with the results of nebular phase spectral modeling studies. \cite{2018Mazzali} find that fitting the nebular phase spectra of SN~2007on and SN~2011iv requires two-component models, consistent with the bimodal identification we obtain for each event. No secondary component is required when modelling SN~2011fe \citep{2015Mazzali}, SN~1991T \citep{2014Sasdelli}, or SN~2004eo \citep{2008Mazzali}, as expected for events we classify as single-component.

Near-infrared (NIR) observations are also consistent with our classification scheme. We identify SN~2003hv as a bimodal event, and flat-topped profiles of the [Fe II] 1.257 and 1.644 $\mu$m features indicate that it is indeed an asymmetric explosion \citep{2006Motohara}. Meanwhile, there are no such indications in NIR observations of SN~2014J, which we classify as a single-component event \citep{2018Dhawan}.

In addition to the six bimodal events we confidently identify through multi-feature fitting, we identify two events where the [Co~III] $\lambda5891$ feature is consistent with bimodality even though the best multi-feature fit does not satisfy our classification criterion. The spectra of these two events -- SNe 2016iuh and 2002er -- are shown in Figure~\ref{fig:CoIIIFit}. Although we regard these identifications as less robust than those obtained through the multi-feature fits, they are nevertheless meaningful and we find no obvious problems when we examine the spectra manually. Particularly in the case of SN~2002er, the structure of the [Co~III] feature is clearly double-peaked.
The [Co~III] feature could be affected by Na~I~D absorption. However, its color excess of $E(B-V)=0.36$ corresponds to an Na~I~D equivalent width of $\sim1.2${\AA} \citep{2004Pignata,2012Poznanski} that is nearly three times smaller than the absorption feature observed in the spectrum. While \cite{2013Phillips} showed that there is significant dispersion in this relation, it is unlikely that that the double peaks in this feature can be explained as a product of host extinction.
Furthermore, we find that the peak absolute magnitude of these two events are very similar to those identified through multi-feature fitting.

SN~1986G presents a unique challenge and warrants further discussion. Our best fit convolution model to this spectrum is shown in Figure~\ref{fig:Template} and is clearly consistent with a bimodal identification under the classification scheme described above. However, this event suffers from considerable host galaxy extinction, and the presence of absorption from the Na~I~D doublet ($\lambda\lambda5890,5896$) complicates interpretation of the [Co~III] $\lambda5891$ emission feature. There is also some concern that a portion of the central minimum of the double-peaked $\sim6600$~{\AA} [Co~III]/[Fe~II] feature may be an artifact introduced through over-subtraction of the host galaxy contribution.

While these complications may produce artificial absorption in the central regions of the emission features in question, they will not affect the wings (which are well-matched by the best-fit model, particularly so for the [Co~III] $\lambda5891$ and $\sim5300$~{\AA} [Fe~II]/[Fe~III] features). In fact the artificial absorption may help explain why the two features in question have deeper central minima than produced by the otherwise well-fitting convolution. Nevertheless, due to the difficulty of interpretation we regard this event as a tentative bimodal identification. It is denoted by orange markers in subsequent figures.

\begin{table}
\caption{Convolution Fit Parameters.}
\label{tab:fitparams}
\begin{tabular}{lrrrrl}
\hline\hline
SN & $v_{shift}$ & $\sigma_{mod,1}$ & $\sigma_{mod,2}$ &  $v_{sep}$ & $r$ \\
\hline\hline
SN 2014bv & $-$1007 & 847 & 3130 & 5154 & 0.441 \\
SN 2011iv & $-$2327 & 1868 & 4163 & 6198 & 0.740 \\
SN 2007on & $-$2170 & 1161 & 1799 & 5675 & 0.603 \\
SN 2003gs & 1300 & 425 & 373 & 4321 & 0.768 \\
SN 2003hv & $-$2996 & 1211 & 5116 & 6460 & 0.636  \\
SN 2005am & $-$1036 & 3083 & 3837 & 6999 & 0.858  \\
SN 2016iuh & $-$2373 & 1307 & 3777 & 4776 & 0.658 \\
SN 2002er & $-$926 & 3579 & 2770 & 6074 & 1.056  \\
SN 1986G & $-$2255 & 1065 & 2178 & 4175 & 0.729 \\
SN 2012cg & $-$3238 & 426 & 8368 & 6582 & 0.698  \\
SN 2012ei & $-$592 & 1185 & 5581 & 5026 & 0.838 \\
\hline\hline
\end{tabular} \\
\begin{flushleft}
These fit parameters are the same as those described in Section~\ref{sec:bimodality}.\\
Note that velocities are presented in units of km s$^{-1}$.
\end{flushleft}
\end{table}

\section{Discussion and Conclusions}
\label{sec:conclusion}

\begin{figure}
\centering
\includegraphics[scale=0.51]{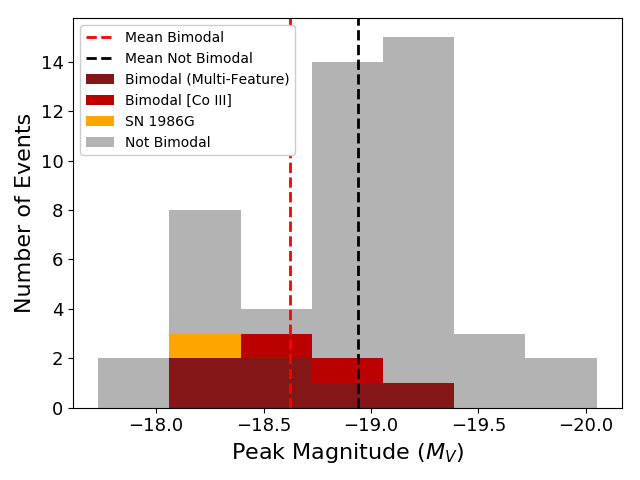}
\caption{The distribution of peak $M_V$ for our sample of SNe~Ia. Spectra showing evidence of bimodal $^{56}$Ni velocity profiles are shown in red, and those without such evidence are shown in grey. The different shades of red indicate whether a spectrum was identified as bimodal using fits to all three $^{56}$Ni decay features (darker) or based only upon fits to the [Co~III] $\lambda5891$ feature (lighter). Note that the average peak $M_V$ for those events showing evidence of bimodal $^{56}$Ni velocity profiles is fainter than those without such evidence by 0.32 mag.}
\label{fig:M_VDist}
\end{figure}

\begin{figure*}
\includegraphics[width=0.49\textwidth]{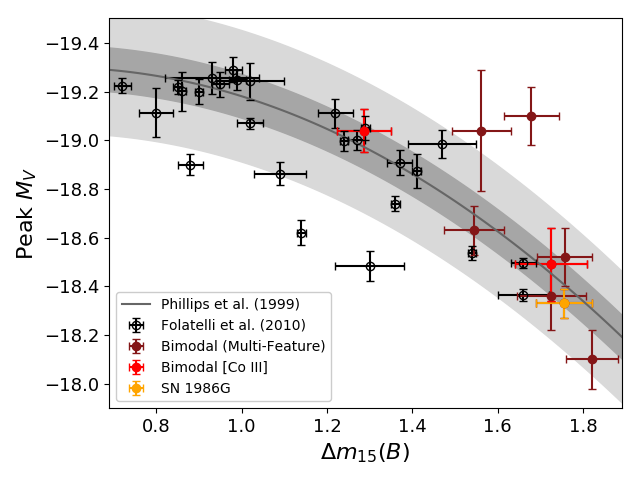}
\hfill
\includegraphics[width=0.49\textwidth]{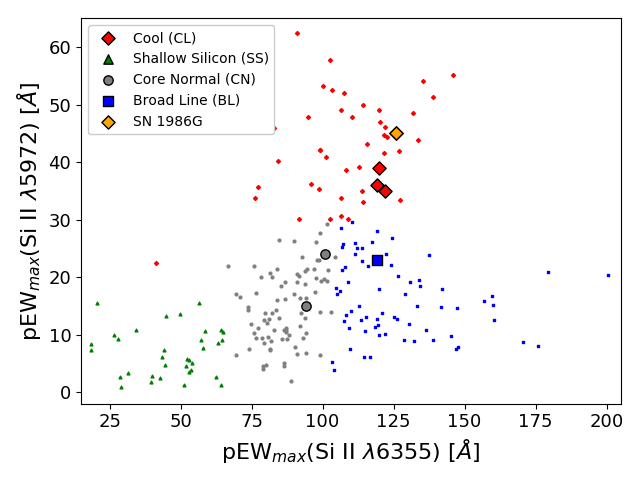}
\caption{Empirical comparisons of the bimodal SNe~Ia in our sample with the SNe~Ia population at large. The left panel shows the decline rate versus peak luminosity for bimodal events (shown in red) compared to the \protect\cite{1999Phillips} relation. A sample of SNe~Ia from \protect\cite{2010FolatelliCSP} are also included for comparison (shown in black). Events identified as bimodal using fits to all three $^{56}$Ni decay features are shown using darker red markers, while identifications based only upon fits to the [Co~III] $\lambda5891$ feature are shown with lighter red markers. The right panel shows the pseudo-equivalent width (pEW) values for the Si~II $\lambda5972$ and Si~II $\lambda6355$ absorption features in near maximum light spectra for six of the bimodal events (shown using large symbols). The different symbols and colors correspond to the different subclasses defined by \protect\cite{2006Branch}. Formal pEW measurement uncertainties are of order 1 {\AA} and are comparable in size to the symbols used to plot the bimodal events. We include the SNe~Ia sample from \protect\cite{2012Blondin} in this Branch diagram for comparison (shown using smaller symbols). Aside from their tendency toward lower peak luminosities, the bimodal SNe~Ia do not appear to be significant outliers in either distribution.}
\label{fig:EmpiricalComparisons}
\end{figure*}

Figure~\ref{fig:M_VDist} shows the peak absolute $V$-band magnitude $(M_V)$ distribution for all of the SNe~Ia in our sample. This distribution is similar to the SNe~Ia luminosity distribution found by \cite{2016Ashall} when neglecting host galaxy extinction. Events shown in gray are consistent with an underlying single-component velocity profile, while those shown in red are identified as bimodal utilizing the classification scheme described in Section~\ref{sec:bimodality}. The darker shade of red indicates an event confidently identified as bimodal using the multi-feature fit criteria, while the lighter shade indicates an event whose [Co~III] feature is consistent with bimodality.

The most striking characteristic of the bimodal events is their marked tendency toward fainter peak magnitudes. As noted by \cite{2015Dong} in their more limited sample, SNe~Ia showing bimodal velocity profiles tend to have relatively large $\Delta m_{15}(B)$ values and be less luminous than those which do not. We find that the average peak $M_V$ for events without signatures of bimodality is $-18.94$ mag (the vertical black dashed line in Figure~\ref{fig:M_VDist}), while that of events with signatures of bimodality is $-18.62$ mag (the vertical red dashed line in Figure~\ref{fig:M_VDist}), a statistically significant offset of $0.32$ mag. Using Welch's t-test we can conclude with 95.7\% confidence that the means of these two distributions are different ($t=2.199; p=0.043$), and using the two-sample Kolmogorov-Smirnov test we can conclude with 95.9\% confidence that the two distributions are distinct from one another ($D=0.487; p=0.041$).

Figure~\ref{fig:EmpiricalComparisons} shows empirical comparisons of the bimodal sample relative to the rest of the SNe~Ia population. The \cite{1999Phillips} decline rate versus peak luminosity relation we show in the figure is calibrated using the $M_{V,peak}[\Delta m_{15}(B) = 1.1]=-19.12$ value from \protect\cite{2010FolatelliCSP}, and we obtained the $\Delta m_{15}(B)$ values for our sample using SuperNovae in Object Oriented Python \citep[SNooPy;][]{2011Burns,2014Burns}. We find that, when compared to the Phillips relation and a sub-sample of SNe~Ia observed by the Carnegie Supernova Project, the bimodal SNe~Ia are not significant outliers. Although they are systematically less luminous at maximum light, they still lie on the Phillips relation and do not show significantly more variance in absolute magnitude at fixed $\Delta m_{15}$ than other SNe~Ia.

We draw a similar conclusion when we examine the bimodal sample using the near-maximum spectroscopic subclasses identified by \cite{2006Branch}, shown in the right panel of Figure~\ref{fig:EmpiricalComparisons} along with a large sample of SNe~Ia from \cite{2012Blondin}. The six bimodal SNe~Ia for which there are publicly available near-maximum spectra appear consistent with the SNe~Ia population at large. They do not fall into a limited range of the Branch diagram parameter space and are reasonably split among the four empirical classifications. These empirical characteristics of the bimodal sample are summarized in Table~\ref{tab:bimodalcharacteristics}.

The tendency toward lower peak luminosities is consistent with observing collisional events at viewing angles of $\theta \sim 90\degree$, perpendicular to the collision axis $(\theta = 0\degree)$. Using the Lagrangian hydrodynamics code of \cite{2008RosswogSPH} and \cite{2009RosswogSPH} and the three-dimensional radiative transfer code SEDONA \citep{2006SEDONA}, \cite{2009Rosswog} simulated explosions and calculated synthetic light curves for WD-WD collisional detonations of varying WD masses. These simulated light curves are similar to those observed for SNe~Ia. \cite{2009Rosswog} also noted that, due to the asymmetry of the resultant ejecta, the observed properties of these SNe would exhibit some degree of viewing angle dependence. They found that the synthetic peak $M_B$ could be reduced by as much as $\sim0.5$ mag when viewed edge on at $\theta \sim 90\degree$. This effect is similar in scale to that which we observe for this sample, although the synthetic light curves calculated by \cite{2009Rosswog} have somewhat smaller $\Delta m_{15}(B)$ than the bimodal events in our sample.

Another potential physical explanation is that explosions produced through collisions of average mass WDs may synthesize less $^{56}$Ni than other progenitor channels. \cite{2013Kushnir} calculated the $^{56}$Ni yields $(M_{56})$ produced from collisional explosions for a range of binary WD masses. Their simulations showed that collisions between WDs of mass $0.55-0.65~\text{M}_\odot$ produce between $0.2$ and $0.4~\text{M}_\odot$ of $^{56}$Ni. This mass regime is important because the WD mass function is strongly peaked at $\sim 0.6~\text{M}_\odot$ \citep{2007Kepler}. As shown by \cite{2014Piro}, the combination of the WD mass function and the collisional model of \cite{2013Kushnir} predicts a $^{56}$Ni yield distribution peaked near $0.3~\text{M}_\odot$. Because the distribution of $^{56}$Ni yields inferred from observed SNe~Ia peaks at $M_{56} = 0.6~\text{M}_\odot$ with an average value of $\sim 0.5~\text{M}_\odot$ \citep{2006Stritzinger,2008Wang}, and because the peak Ia luminosity is directly connected with the synthesized $^{56}$Ni mass, \cite{2014Piro} concluded that SNe~Ia produced by collisions should be sub-luminous.

The expected scale of this effect can be estimated using the decline rate-nickel mass relation of \cite{2007Mazzali},
\begin{equation}
    M_{56}/\text{M}_\odot = 1.34 - 0.67 \Delta m_{15}(B),
\end{equation}
and the $V$-band peak luminosity fit from \cite{1993Phillips},
\begin{equation}
    M_{V,max} = -20.883+1.949\Delta m_{15}(B).
\end{equation}
If the collisional WD channel produces SNe~Ia with $M_{56}\sim0.3~\text{M}_\odot$, we would thus expect them to be $\sim0.6$ mag fainter than typical SNe~Ia (with $M_{56}\sim0.5~\text{M}_\odot$), which is comparable to the effect observed in our sample.

We can estimate the critical viewing angle ($\theta_c$, measured relative to the WD collision axis) beyond which we are no longer sensitive to identifying bimodal signatures. This critical angle is given by $\theta_c = 90\degree - \cos^{-1}{ (P_{bimodal}/f) }$, where $P_{bimodal}$ is the proportion of observed SNe~Ia for which we detect signatures of bimodality, and $f$ is the fraction of all SNe~Ia with intrinsic bimodal velocity components. Thus, if we assume that all SNe~Ia have intrinsic bimodality, then we are sensitive to viewing angles up to $\theta_c \sim 10\degree$. If we assume instead that $f=1/3$, then we are sensitive to viewing angles up to $\theta_c \sim 30\degree$. It seems unlikely that we would be sensitive to $\theta_c \gtrsim 45\degree$, so the fraction of SNe~Ia with intrinsic bimodalility is probably not significantly smaller than 25\%.

A third potential explanation is that the method we use in this paper may simply be biased towards detecting bimodality in fainter SNe~Ia. Abundance tomography studies indicate that $^{56}$Ni is distributed out to considerably higher velocities in normal SNe~Ia when compared to subluminous events. In the normal SN~Ia 2011fe, $^{56}$Ni is inferred to extend to velocities beyond 10,000 km s$^{-1}$ \citep{2015Mazzali}, for example, while in the subluminous SN~Ia 1986G it is inferred to extend only to about 6,000 km s$^{-1}$ \citep{2016AshallB}. In order to observe signatures of bimodality, the two WDs need to collide with a velocity comparable to that of the $^{56}$Ni region, so slower moving $^{56}$Ni distributions would be more readily detectable.

However, this potential bias seems unlikely to be a significant effect. If this were a dominant effect one would expect to find numerous events with slightly overlapping velocity components that satisfy our single [Co~III] feature fit criteria, and one would expect those events to be considerably brighter than the rest of the bimodal sample. We observe neither of those outcomes. It is important to note that abundance tomography studies generally assume a single-component model. When analyzing a spectrum that is comprised of two components, such an assumption would infer $^{56}$Ni distributions extending to artificially high velocities. This can be seen clearly in the case of SN~2007on. When using a single-component model, $^{56}$Ni in the ejecta is inferred to extend to velocities beyond 12,500 km s$^{-1}$ \citep{Ashall2018}. The presence of $^{56}$Ni at such high velocities would smear out any signature of bimodality, and yet we can clearly identify SN~2007on as a bimodal event (See Figure~\ref{fig:MultiFit}). Detailed modeling by \cite{2018Mazzali} further confirms that the event is better reproduced using a model with two narrow velocity components instead of a single broad component.

It is now established that a non-negligible fraction of SNe~Ia spectra exhibit features consistent with a bimodal $^{56}$Ni velocity distribution. \cite{2015Dong} found that 3 out of the 18 spectra they examined showed compelling evidence of bimodality. Here we more than double the sample and find that 8 out of 47 spectra show evidence of bimodality. The collisional WD scenario provides a possible explanation for these observed spectral properties, and the tendency towards fainter peak luminosities that we report here is also consistent with this theoretical picture. Further improvements will require larger statistical samples and more attention to possible selection effects as the statistical uncertainties on the bimodal fraction become smaller. Nevertheless, we can confidently assert that bimodal events are not rare, and any proposed combination of SNe~Ia explosion scenarios must be able to produce a non-negligible fraction of them.

\begin{table}
\caption{Empirical Characteristics of the Bimodal Sample.}
\label{tab:bimodalcharacteristics}
\begin{tabular}{lccc}
\hline\hline
Name & Peak $M_V$ & $\Delta m_{15}(B)$ & Branch Class \\
\hline\hline
SN 2014bv & $-18.36\pm0.14$ & $1.726\pm0.080$ & ---  \\
SN 2011iv & $-19.10\pm0.12$ & $1.679\pm0.064$ & CL \\
SN 2007on & $-18.52\pm0.12$ & $1.757\pm0.064$ & CL \\
SN 2003gs & $-18.10\pm0.12$ & $1.820\pm0.061$ & ---  \\
SN 2003hv & $-19.04\pm0.25$ & $1.562\pm0.070$ & CN   \\
SN 2005am & $-18.63\pm0.10$ & $1.544\pm0.070$ & BL   \\
SN 2016iuh & $-18.49\pm0.15$ & $1.725\pm0.084$ & CL  \\
SN 2002er & $-19.04\pm0.09$ & $1.286\pm0.063$ & BL  \\
SN 1986G & $-18.33\pm0.06$ & $1.756\pm0.066$ & CL \\
\hline\hline
\end{tabular} \\
\begin{flushleft}
Branch classes are not reported for SNe 2014bv and 2003gs because there are no publicly available early phase spectra for these events.
\end{flushleft}
\end{table}

\section*{Acknowledgments}

We thank the referee for helpful comments. All of the previously unpublished spectra presented in this paper were obtained with the MODS spectrographs built with funding from NSF grant AST-9987045 and the NSF Telescope System Instrumentation Program, with additional funds from the Ohio Board of Regents and the Ohio State University Office of Research. The LBT is an international collaboration among institutions in the United States, Italy and Germany. LBT Corporation partners are: The Ohio State University, and The Research Corporation, on behalf of The University of Notre Dame, University of Minnesota and University of Virginia; The University of Arizona on behalf of the Arizona university system; Istituto Nazionale di Astrofisica, Italy; LBT Beteiligungsgesellschaft, Germany, representing the Max-Planck Society, the Astrophysical Institute Potsdam, and Heidelberg University.

We thank Christa Gall for providing the data for SN 2011iv, and we are grateful to Todd Thompson for valuable comments. We thank the Las Cumbres Observatory and its staff for its continuing support of the ASAS-SN project. We thank Subo Dong and Ping Chen for providing us with reduced SNIPER photometry from LCOGT data for a number of events. ASAS-SN is supported by the Gordon and Betty Moore Foundation through grant GBMF5490 to the Ohio State University and NSF grant AST-1515927. Development of ASAS-SN has been supported by NSF grant AST-0908816, the Mt. Cuba Astronomical Foundation, the Center for Cosmology and AstroParticle Physics at the Ohio State University, the Chinese Academy of Sciences South America Center for Astronomy (CAS- SACA), the Villum Foundation, and George Skestos.
 
PJV is supported by the National Science Foundation Graduate Research Fellowship Program Under Grant No. DGE-1343012. MAT acknowledges support from the United States Department of Energy through the Computational Sciences Graduate Fellowship (DOE CSGF). KZS and CSK are supported by NSF grants AST-1515876, AST-1515927, and AST-1814440.

This research has made use of the NASA/IPAC Extragalactic Database (NED), which is operated by the Jet Propulsion Laboratory, California Institute of Technology, under contract with the National Aeronautics and Space Administration. This research has made use of NASA's Astrophysics Data System Bibliographic Services. IRAF is distributed by the National Optical Astronomy Observatory, which is operated by the Association of Universities for Research in Astronomy (AURA) under a cooperative agreement with the National Science Foundation.


\label{lastpage}

\end{document}